\documentclass[conference]{IEEEtran}

\usepackage{cite}
\usepackage{amsmath,amssymb,amsfonts}
\usepackage{algorithmic}
\usepackage{graphicx}
\usepackage{textcomp}
\usepackage[table]{xcolor}

\def\BibTeX{{\rm B\kern-.05em{\sc i\kern-.025em b}\kern-.08em
    T\kern-.1667em\lower.7ex\hbox{E}\kern-.125emX}}

\usepackage{ctable}
\usepackage{enumitem}
\usepackage{url}

\newcommand{\tweakedsim}{\raise.17ex\hbox{$\scriptstyle\mathtt{\sim}$}}

\definecolor{MyGold}{HTML}{E69F00}
\definecolor{MyOrange}{HTML}{D55E00}
\definecolor{MyRed}{HTML}{CF0000}
\definecolor{MyGreen1}{HTML}{009E73}
\definecolor{MyGreen}{HTML}{0BC681}
\definecolor{MyBlue1}{HTML}{0072B2}
\definecolor{MyBlue}{HTML}{0E1589}
\definecolor{MyGray}{gray}{0.8}
\definecolor{MyBlack}{HTML}{000000}

\begin{document}

\title{Analytics of Longitudinal System Monitoring Data for Performance
Prediction}

\author{\IEEEauthorblockN{Ian J.~Costello}
\IEEEauthorblockA{
  Google, Inc.\\
  Mountain View, USA\\~\\
  ianjc@google.com
}
\and
\IEEEauthorblockN{Abhinav Bhatele}
\IEEEauthorblockA{
  Department of Computer Science,\\University of Maryland\\
  College Park, USA\\
  bhatele@cs.umd.edu
}
}

\maketitle

\begin{abstract}
In recent years, several HPC facilities have started continuous monitoring of
their systems and jobs to collect performance-related data for understanding
performance and operational efficiency. Such data can be used to optimize the
performance of individual jobs and the overall system by creating data-driven
models that can predict the performance of jobs waiting in the scheduler queue.
In this paper, we model the performance of representative control jobs using
longitudinal system-wide monitoring data and machine learning to explore the
causes of performance variability. We analyze these prediction models in great
detail to identify the features that are dominant predictors of performance. We
demonstrate that such models can be application-agnostic and can be used for
predicting performance of applications that are not included in training. 

\end{abstract}


\section{Motivation}
\begin{figure}[h]
\centering
\includegraphics[width=\columnwidth]{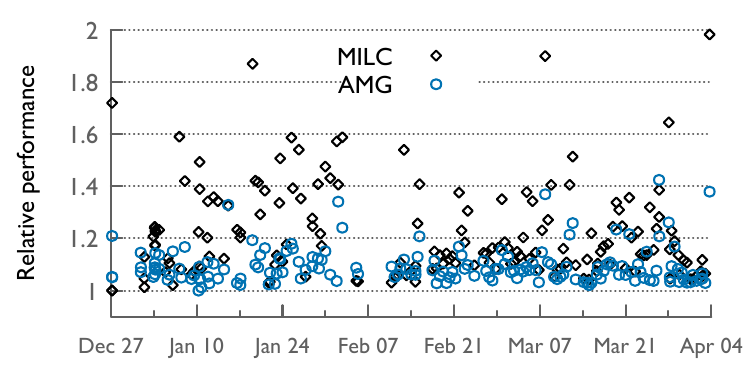}
\caption{Variability in the performance of 128-node AMG and MILC jobs on
different days (on Cori at NERSC).}
\label{fig:variability}
\end{figure}

Run-to-run variability in the performance of parallel codes running on
production high performance computing (HPC) platforms is a real
problem~\cite{PeKePa03, bhatele:sc2013, bhatele:ipdps2020}.
Figure~\ref{fig:variability} shows the varying performance of two HPC
applications, Algebraic Multigrid (AMG) and MIMD Lattice Computation (MILC), in
spite of running the same executable and input in each job. This variability
occurs either due to operating system noise impacting compute regions in the
code or due to varying load on shared resources such as the network or
filesystem because of changing workloads on the system. There are several ways
to mitigate the former but diagnosing and mitigating the impact of the latter
is still a challenge on many HPC systems.

Sharing of network or filesystem resources by all concurrently running jobs
leads to uneven resource usage over time, which can impact performance
reproducibility. For individual HPC users, reducing performance variability
leads to more predictable performance, faster scientific results, and reduced
allocation costs. At a system level, better performance of individual jobs
leads to both energy savings and higher overall job throughout.

In recent years, several HPC facilities have started continuous monitoring of
their systems and jobs to collect performance-related data for understanding
performance and operational efficiency~\cite{lockwood:sc2018}.  Analyzing such
data and using it for data-driven modeling can facilitate understanding the
causes of performance variability, forecasting performance, and in providing
insights that can translate into actions for correcting inefficient behavior.
For example, an intelligent software stack can use modeling to predict the
runtime of jobs waiting in the scheduler queue and use those models to adapt
scheduling decisions to mitigate performance
variability~\cite{nichols:ipdps2022}.

In this paper, we analyze longitudinal system monitoring data to explore the
causes of performance variability in certain {\em control} jobs. The monitoring
data is collected by the Lightweight Distributed Metric Service (LDMS)~\cite{ldms, agelastos:parco2016} on Cori,
a \tweakedsim 30 Pflop/s Cray XC40 system (recently retired) at NERSC.
Analyzing system-wide monitoring data gives us a global view of the system, a
perspective which users running individual jobs do not have. Separately, we
also run some control jobs on Cori to document the impact of varying resource
usage on application performance. Our goal is to use information about the
system state before a job starts executing to create a model that can predict
the performance of future jobs based on the system state at that time.

LDMS is becoming more prevalent on HPC systems, specially at Department of
Energy centers and even some NSF centers. The Blue Waters machine at NCSA has
used LDMS for a long time. ALCF/ANL, LC/LLNL, and NERSC/LBL are all using LDMS
for collecting monitoring data on their clusters. While specific
metrics/features may be different depending on the compute and networking
hardware, the methodology developed in this paper should translate without much
effort to other centers running LDMS.

We use machine learning, specifically regression models, to model the execution
time of jobs in terms of several network related hardware counters gathered by
LDMS. We use these models to understand which hardware counters are strong
predictors of performance, in other words, indicators of performance
degradation. We demonstrate that our data-driven modeling approach that uses
past system state is successful in performance prediction of unseen data --
i.e. new jobs waiting in the scheduler queue.

\vspace{0.1in}
\noindent Our work makes the following important contributions:
\begin{itemize}[leftmargin=*]
\item We create a pipeline to process, filter and aggregate large-scale
system-wide monitoring data to make it suitable for consumption by ML models.
\item We develop ML-based regression models that can predict the performance of
unseen jobs using past system state.
\item We analyze feature importances in different models to identify strong
predictors of performance degradation and show how lightweight monitoring of
these counters can lead to powerful insights prior to job execution.
\item We demonstrate that such models can be application-agnostic and can be
used for predicting performance of applications that are not included in the
training data.
\end{itemize}

\section{Background and Related Work}
\begin{figure}[h]
\centering
\includegraphics[width=\columnwidth]{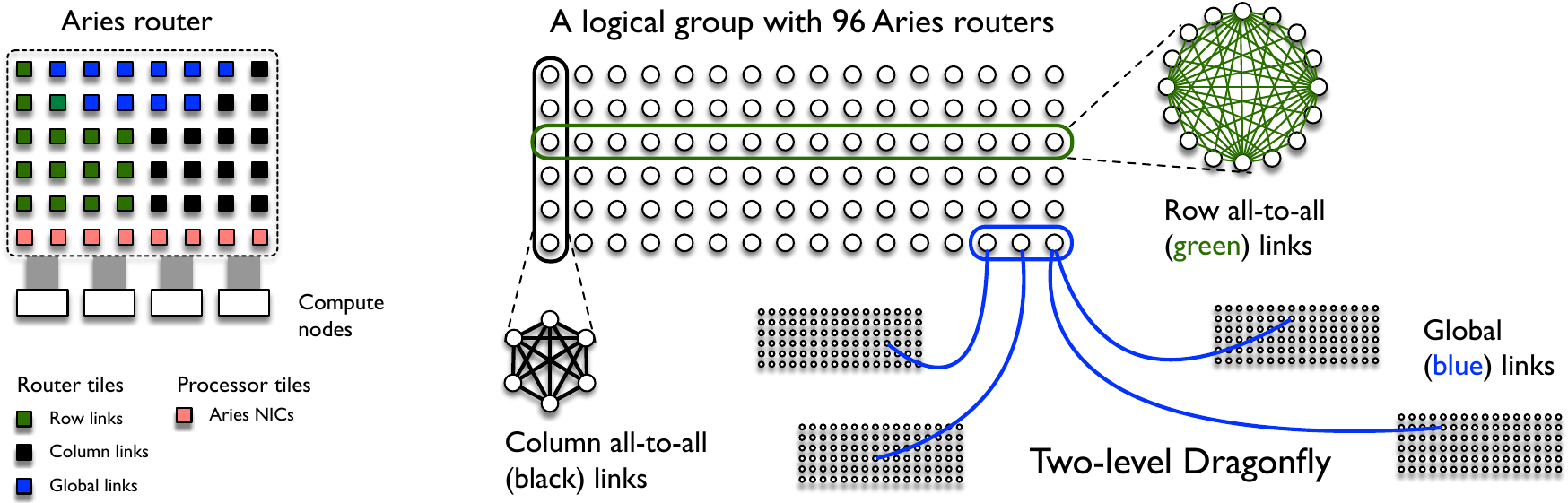}
\caption{Network ports classified into router and processor tiles on the 48-port
Aries router (left) and a multi-group dragonfly system constructed using the
Aries router as a building block (right).}
\label{fig:dragonfly}
\end{figure}

In this section, we provide background on sources of performance variability
and related work.  We first briefly describe the dragonfly network deployed in
Cori (Cray XC40), which is used for experiments in this paper.  The Cray XC40
system uses the Aries router to create a dragonfly network
topology~\cite{kim:dragonfly}. The Aries router has 48 ports that are used to
connect to compute nodes and other routers on the network (see
Figure~\ref{fig:dragonfly}). Eight ports referred to as processor tiles are
used to connect to four compute nodes. The remaining 40 ports are referred to
as router tiles and are used to connect to other routers. 96 Aries routers are
connected in a 16 $\times$ 6 rectangular grid to form a logical group. In a
group, each router is directly connected to all other routers in its row and
all other routers in its column. The remaining ports are used for global links
which connect to routers in other groups throughout the system. 

\subsection{Resource Management and Sources of Variability}

On most HPC systems, compute nodes are dedicated to individual jobs but the
global network and filesystem are shared by all concurrently running jobs.
Moreover, job schedulers assign any available nodes to ``ready'' jobs in the
queue without regard to the nodes' location in the physical network topology.
Cray dragonfly-based systems are no exception to this.
There are no guarantees provided by the job scheduler as to the ``compactness''
of a job allocation. Hence, a job may share routers and groups with other jobs. 

In principle, the compactness of a job should have no bearing on its
communication or overall runtime because adaptive indirect routing deployed on
dragonfly networks should distribute traffic evenly over all global
links~\cite{ugal}. However, in practice, significant performance variability
can be observed when the same executable is run on a dragonfly system
repeatedly. This paper primarily considers such effects of network traffic on
variability. Traffic on the network consists of within-job communication, for
example MPI messages, and traffic from/to the filesystem during input/output
(I/O) operations. Even in the absence of OS noise, this resource sharing by
concurrently running jobs can have a significant performance impact. 

\begin{table*}[t]
\centering
\caption{Description of raw network hardware performance counters gathered by LDMS. These raw counters are aggregated within a router across network tiles or processor tiles.}
\begin{tabular}{lll} \toprule
& Raw counter name & Description \\ \midrule
\cellcolor{MyGold}   & {\tt AR\_RTR\_INQ\_PRF\_INCOMING\_FLIT\_VCv}        & Number of flits received on virtual channel v of a router tile (8 counters)\\
\cellcolor{MyOrange} & {\tt AR\_RTR\_INQ\_PRF\_INCOMING\_PKT\_VCv}         & Number of cycles stalled on a virtual channel v of a router tile (8 counters) \\
\cellcolor{MyRed}    & {\tt AR\_RTR\_INQ\_PRF\_ROWBUS\_STALL\_CNT}         & Total number of cycles stalled on a router tile \\ \midrule
\cellcolor{MyGreen1} & {\tt AR\_NL\_PRF\_REQ\_NIC\_n\_TO\_PTILES\_FLITS}   & Number of NIC request flits from NIC n to all processor tiles \\
\cellcolor{MyGreen1} & {\tt AR\_NL\_PRF\_REQ\_PTILES\_TO\_NIC\_n\_FLITS}   & Number of NIC request flits from all processor tiles to NIC n\\
\cellcolor{MyBlue1}  & {\tt AR\_NL\_PRF\_RSP\_NIC\_n\_TO\_PTILES\_FLITS}   & Number of NIC response flits from NIC n to all processor tiles \\
\cellcolor{MyBlue1}  & {\tt AR\_NL\_PRF\_RSP\_PTILES\_TO\_NIC\_n\_FLITS}   & Number of NIC response flits from all processor tiles to NIC n \\
\cellcolor{MyGray}   & {\tt AR\_NL\_PRF\_REQ\_NIC\_n\_TO\_PTILES\_STALLED} & Number of clock cycles requests from NIC n have stalled to all processor tiles \\
\cellcolor{MyGray}   & {\tt AR\_NL\_PRF\_REQ\_PTILES\_TO\_NIC\_n\_STALLED} & Number of clock cycles requests from all processor tiles have stalled to NIC n \\
\cellcolor{MyBlack}  & {\tt AR\_NL\_PRF\_RSP\_NIC\_n\_TO\_PTILES\_STALLED} & Number of clock cycles responses from NIC n have stalled to all processor tiles \\
\cellcolor{MyBlack}  & {\tt AR\_NL\_PRF\_RSP\_PTILES\_TO\_NIC\_n\_STALLED} & Number of clock cycles responses from all processor tiles have stalled to NIC n \\ \bottomrule
\end{tabular}
\label{tab:counters-raw}
\end{table*}

\subsection{Related Work}

Performance variability refers to the variation in the execution time of an
application executable with the same input parameters across multiple
executions.  Several studies have established the significant differences in
performance between identical executions of the same HPC job~\cite{PeKePa03,
hoefler:sc2010, bhatele:sc2013, groves:perf_variability_on_dragonfly,
bhatele:ipdps2020}. Petrini et al.~\cite{PeKePa03} highlight the role of
operating system daemons in creating noise and in degrading application
performance.

Research on collecting system data for analysis and forecasting is not new.
More than 25 years ago, Wolski et al.~\cite{wolski:sc97} developed a system
called the Network Weather Service (NWS) to estimate CPU usage and throughput
of network traffic based on system state and statistical models. In 2005,
Skinner et al.~\cite{skinner2005understanding} highlighted the impact of
cross-application contention and parallel filesystem interference on the NERSC
IBM SP system (Seaborg). 

In recent years, HPC researchers have begun to explore the breadth of
historical data available due to comprehensive data gathering at HPC
facilities~\cite{lockwood:sc2018, tuncer:tpds2019}.  Lockwood et
al.~\cite{lockwood:sc2018} discuss performance variations due to varying load
on the I/O sub-system.  Tuncer et al.~\cite{tuncer:tpds2019} perform
classification and detection of anomalous performance based on Aries counters.
They use machine learning combined with system data, but primarily focus on
diagnosing anomalies in compute node health and not performance of jobs. Hoppe
et al.~\cite{hoppe:2020} demonstrate a data pipeline for LDMS to classify
anomalies on a smaller scale than the system that we use in this paper (52
versus 12,076 compute nodes).  Agelastos et al.~\cite{agelastos:parco2016}
create a HPC system profiler to explain the performance variability of
applications across different HPC systems.

Jha et al.~\cite{jha:cluster2018} investigate the relationship between overall
during-run network congestion and performance.  Chunduri et
al.~\cite{Chunduri:2017} measure and attribute runtime variation to runtime
system state. However, they primarily focus on single node variation using data
from the same time period as the execution of a job.
In~\cite{chunduri:pmbs2019}, Chunduri et al.~perform a thorough analysis and
build regression models of job performance using network counters from during
the run. Bhatele et al.~\cite{bhatele:ipdps2020} analyze data from during a
job's execution and use regression models to analyze per-job data combined with
system data to predict the overall performance and per time step performance of
individual jobs. However, they use data from when a job is running and in this
work, we use only system data strictly before a job's execution. Nichols et
al.~\cite{nichols:ipdps2022} train machine learning models using system
monitoring data to predict the occurrence of variation, and use these models in
a job scheduler to alter scheduling decisions.
Our proposed pipeline and analysis remain relevant across different HPC systems
and routers as our work can be applied to any system with LDMS data collection.

\section{Data Description}
\label{sec:data}
We now describe the system-wide data obtained from LDMS, and control jobs' data
used for training ML models. The data was gathered on Cori, a Cray XC40 system
at NERSC. Cori features 12,076 compute nodes across 34 groups; of these 9,668
nodes are powered by 68-core Intel Xeon Phi Knights Landing (KNL)
processors~\cite{nersc:cori_system}.

\subsection{Longitudinal System Monitoring Data}

LDMS enables system administrators to collect system-wide data on HPC clusters
at a configurable frequency from a few seconds to minutes.
We now describe the subset of the LDMS data that we use in our analysis and the
processing of this data to facilitate ingestion by machine learning models.

\vspace{0.08in}
\noindent{\bf Raw LDMS Data:}
Each Aries router has a multitude of hardware counters that track various
network events across each router and processor tile~\cite{ariescounters}. The
raw LDMS data collected on Cori consists of a subset of these hardware
counters, collected every second for each of the 48 network tiles, across all
2890 routers on the system. In this paper, we consider a subset of these
counters (see Table~\ref{tab:counters-raw}, column 1) that we believe to be important
indicators of network congestion.

\begin{table*}[t]
\centering
\caption{Description of derived counters used for modeling. Colors in the left
column provide a mapping of derived features in this table to raw features in
Table~\ref{tab:counters-raw}.}
\begin{tabular}{llll} \toprule
& Derived counter name & Abbreviation & Description for interpreting counter \\ \midrule
\cellcolor{MyGold}   & {\tt AR\_RTR\_INQ\_PRF\_INCOMING\_FLIT\_REQ} & {\tt RT\_FLIT\_REQ} & Total number of request flits received on a router tile \\
\cellcolor{MyGold}   & {\tt AR\_RTR\_INQ\_PRF\_INCOMING\_FLIT\_RSP} & {\tt RT\_FLIT\_RSP} & Total number of response flits received on a router tile \\
\cellcolor{MyOrange} & {\tt AR\_RTR\_INQ\_PRF\_INCOMING\_PKT\_REQ}  & {\tt RT\_PKT\_REQ}  & Total number of cycles requests stalled on a router tile \\
\cellcolor{MyOrange} & {\tt AR\_RTR\_INQ\_PRF\_INCOMING\_PKT\_RSP}  & {\tt RT\_PKT\_RSP}  & Total number of cycles responses stalled on a router tile \\ \midrule
\cellcolor{MyGold}   & {\tt AR\_RTR\_INQ\_PRF\_INCOMING\_FLIT\_ROW} & {\tt RT\_FLIT\_ROW} & Total number of flits received on all row links of a router\\
\cellcolor{MyGold}   & {\tt AR\_RTR\_INQ\_PRF\_INCOMING\_FLIT\_COL} & {\tt RT\_FLIT\_COL} & Total number of flits received on all column links of a router\\
\cellcolor{MyGold}   & {\tt AR\_RTR\_INQ\_PRF\_INCOMING\_FLIT\_GBL} & {\tt RT\_FLIT\_GBL} & Total number of flits received on all global links of a router\\
\cellcolor{MyRed}    & {\tt AR\_RTR\_INQ\_PRF\_ROWBUS\_STALL\_ROW}  & {\tt RT\_STL\_ROW}  & Total number of stalls on all row links of a router\\
\cellcolor{MyRed}    & {\tt AR\_RTR\_INQ\_PRF\_ROWBUS\_STALL\_COL}  & {\tt RT\_STL\_COL}  & Total number of stalls on all column links of a router\\
\cellcolor{MyRed}    & {\tt AR\_RTR\_INQ\_PRF\_ROWBUS\_STALL\_GBL}  & {\tt RT\_STL\_GBL}  & Total number of stalls on all global links of a router\\ \midrule
\cellcolor{MyGreen1} & {\tt AR\_NL\_PRF\_REQ\_FLITS}                & {\tt PT\_FLIT\_REQ} & Total number of NIC request flits on a processor tile\\
\cellcolor{MyBlue1}  & {\tt AR\_NL\_PRF\_RSP\_FLITS}                & {\tt PT\_FLIT\_RSP} & Total number of NIC response flits on a processor tile \\
\cellcolor{MyGray}   & {\tt AR\_NL\_PRF\_REQ\_STALLED}              & {\tt PT\_STL\_REQ}  & Total number of cycles requests stalled on a processor tile \\
\cellcolor{MyBlack}  & {\tt AR\_NL\_PRF\_RSP\_STALLED}              & {\tt PT\_STL\_RSP}  & Total number of cycles responses stalled on a processor tile \\ \bottomrule
\end{tabular}
\label{tab:counters-derived}
\end{table*}

\begin{figure}[h]
\centering
\includegraphics[width=2in]{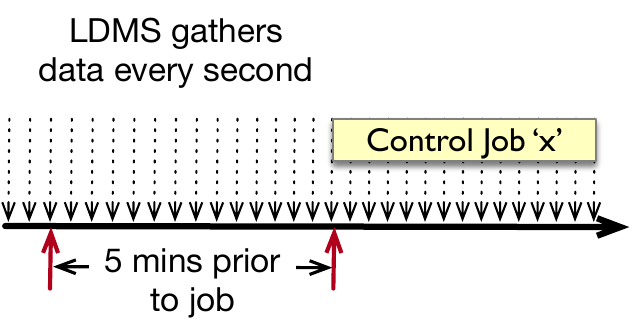}
\caption{LDMS data five minutes prior to job start is used as input to train the machine learning models.}
\label{fig:ldms}
\end{figure}

\vspace{0.08in}
\noindent{\bf Data Extraction from Time Series:}
The raw LDMS counter data is essentially a time series. Since we are interested
in predicting the performance of unseen jobs using system data from the recent
past, we decided to extract data for the last five minutes prior to the
beginning of execution of each control job (described in
Section~\ref{sec:control-jobs}).  For each control job, we filter the time
series by only looking at the last five minutes prior to the start time of a
job and extracting the change in counter values (see Figure~\ref{fig:ldms}).
For each job, this gives us a table of key-value pairs that are the previously
identified network counters and the change in their values in the last five
minutes. This data is available for each router and network tile on the system.

LDMS is becoming more prevalent on HPC systems, specially at Department of
Energy centers and even some NSF centers. The Blue Waters machine at NCSA has
used LDMS for a long time. ALCF/ANL, LC/LLNL, and NERSC/LBL are all using LDMS
for collecting monitoring data on their clusters. While specific
metrics/features may be different depending on the compute and networking
hardware, the methodology developed in this paper should translate without much
effort to other centers running LDMS.

\subsection{Reduction in Data Dimensionality}

The resulting data extracted from the raw time series is still extremely
high-dimensional because the counter values are per network tile (port) and
per router. We perform aggregations of this data along different axes to get
the data in final form:

\vspace{0.08in}
\noindent{\bf Reducing Per Tile Data:}
We begin by performing reductions within each individual router.
Each 48-port Aries router contains 40 router tiles or ports that connect to
other routers and the remaining eight processor tiles or ports connect to the
compute nodes on that router. The aggregation across all router tiles yields
the 17 (2 $\times$ 8 VCs + row\_bus stalls) counters derived from the top
section of Table~\ref{tab:counters-raw}. A similar reduction across all the
processor tiles yields derived counters corresponding to the bottom section of
Table~\ref{tab:counters-raw}.  After this reduction, we are left with $17+8=25$
derived features for each of the 2,890 routers for each job run in our dataset.

\vspace{0.08in}
\noindent{\bf Creating Interpretable Derived Features:} Next, we perform a sum
over different virtual channel groups to create a set of human interpretable
derived features (Table~\ref{tab:counters-derived}).  For example,
\begin{equation}
\begin{split}
\mathtt{AR\_RTR\_INQ\_PRF\_INCOMING\_FLIT\_REQ} = \\
\sum_{v=0}^{3} \mathtt{AR\_RTR\_INQ\_PRF\_INCOMING\_FLIT\_VCv}
\end{split}
\end{equation}
\begin{equation}
\begin{split}
\mathtt{AR\_RTR\_INQ\_PRF\_INCOMING\_FLIT\_RSP} = \\
\sum_{v=4}^{7} \mathtt{AR\_RTR\_INQ\_PRF\_INCOMING\_FLIT\_VCv}
\end{split}
\end{equation}
In the equations above, $\mathtt{v}$ is the virtual channel number that appears
in the column names in the first two rows of Table~\ref{tab:counters-raw}.
There are eight virtual channels and $\mathtt{v}$ iterates over the virtual
channel ID (0 to 3 in Equation 1, and 4 to 7 in Equation 2).

Such groupings lead to the top section of Table~\ref{tab:counters-derived}).
Similarly {\tt AR\_NL\_PRF\_REQ\_NIC\_n\_TO\_PTILES\_FLITS} and {\tt
AR\_NL\_PRF\_REQ\_PTILES\_TO\_NIC\_n\_FLITS} for a processor tile are summed
together to create the {\tt AR\_NL\_PRF\_REQ\_FLITS} feature (bottom section of
Table~\ref{tab:counters-derived}). The colors in Table~\ref{tab:counters-raw}
and~\ref{tab:counters-derived} provide a mapping between the raw and derived
features. The middle section of Table~\ref{tab:counters-derived} represents
another way of creating derived features.  Instead of reducing counters over
all router tiles, we reduce the flit and stall counters by the type of port or
link (row, column, or global). This creates the six features in the middle
section of Table~\ref{tab:counters-derived}.  These different derivations yield
14 derived features for each of the 2,890 routers.

\vspace{0.08in}
\noindent{\bf Filtering by Router Type:}
Different types of nodes such as compute nodes, I/O servers, management and
login nodes are attached to different routers. We can either consider all
routers or filter the data by the types of nodes attached to a router. We
explored the following groupings, each of which only considers a subset of
routers: routers connected to compute nodes (henceforth referred to as {\em all
routers}), routers connected to I/O servers ({\em IO routers}), and routers
attached to nodes that are assigned to the control job (henceforth referred to
as {\em my routers}). These three groupings yielded the strongest results in
our experiments, and are solutions that can be implemented by system
administrators and individual users respectively.

\vspace{0.08in}
\noindent{\bf Aggregating over Routers:}
For each grouping described above, we explored various schemes to aggregate the
data across routers for the set of derived features. This aggregation
calculates one value for each feature by applying one of the following
functions over all the routers that belong to a group: mean, standard
deviation, various percentiles such as median, 75th percentile, 95th
percentile, and IQR (75th -- 25th percentile). We present results for some of
these aggregation functions in Section~\ref{sec:results}.

\subsection{Control Jobs Data}
\label{sec:control-jobs}

We set up some control jobs that enable us to assess the impact of system state
on the performance of production applications. We run three codes -- AMG, MILC,
and miniVite, which are representative of common workloads on HPC systems.  AMG
and MILC were run in a weak scaling mode on 128 and 512 nodes. miniVite's input
problem is a fixed-size real world graph, and it was only run on 128 nodes.
Each application run was short, running for between five to ten minutes. We
briefly describe each application below.

\vspace{0.08in}
\noindent{\bf AMG}: is a proxy application for parallel algebraic multigrid and
it uses the Hypre linear solver library~\cite{hypre}. In our setup, AMG runs
AMG-GMRES on a linear system for a three-dimensional input problem with
dimensions $32 \times 32 \times 32$ (per MPI process).

\vspace{0.08in}
\noindent{\bf MILC}: refers to MIMD Lattice Computation, used for numerical
simulations of quantum chromodynamics. The MILC application, su3\_rmd, was
used in these experiments, which performs a 4D stencil on a per process grid of
dimensions $4 \times 4 \times 4 \times 4$.

\vspace{0.08in}
\noindent{\bf miniVite}: is a proxy application for Vite~\cite{vite:ipdps18},
and is representative of graph analytics workloads~\cite{minivite:pmbs18}.  It
performs a single phase of the Louvain classification, which is an algorithm
for community detection in large distributed graphs.
An iterative loop was added to miniVite to repeat its work several times.

\vspace{0.08in}
Data for experiments in this paper was collected by submitting control jobs for
each application to the production batch queue on Cori between December 2018
and April 2019. All control jobs were run on KNL nodes, alongside jobs of other
users on the system. Four out of 68 cores on each node were reserved for OS
daemons to minimize the effects of OS noise on compute regions in the code.
The applications 
did not perform any I/O, to rule out variability due to I/O congestion.
Based on when each job ran, LDMS data was processed to obtain the derived
features for the 5-minute interval prior to the execution of the job
(Figure~\ref{fig:ldms}). We also recorded the execution time for the main
execution loop of each application, which is the variable the ML models try to
predict. Table~\ref{tab:apps} presents the five datasets created for training
the models and the number of samples in each dataset. Note that whenever we
refer to an application dataset in the paper, it refers to the system-wide data
collected from the 5-minute interval prior to the beginning of each job in the
dataset and that job's respective execution time.  The models are trained
solely on the system-wide data and no application-specific data is used for
training.

\begin{table}[h]
\centering
\caption{Datasets based on applications run in the control jobs}
\begin{tabular}{lccc} \toprule
Application  & No.~of Nodes & Number of jobs & Dataset Name \\ \midrule
AMG 1.1      & 128 & 156 & AMG 128\\
AMG 1.1      & 512 & 152 & AMG 512\\
MILC 7.8.0   & 128 & 151 & MILC 128\\
MILC 7.8.0   & 512 & 153 & MILC 512\\
miniVite 1.0 & 128 & 119 & miniVite\\
\bottomrule
\end{tabular}
\label{tab:apps}
\end{table}

\noindent{\bf Job Placement Data:} 
We also calculate two additional job-related features from scheduler queue
logs. The first, {\tt NUM\_ROUTERS}, indicates the total number of unique
routers that a job was assigned nodes on. The second, {\tt NUM\_GROUPS},
indicates how many dragonfly groups these routers were spread across.  These
features indicate the degree of compactness or spread in terms of placement of
each job.

\section{Methods for Data Analytics}
In this section, we present our approach for creating prediction models,
obtaining importance of different features, and evaluating the predictive power
of the trained models.

\subsection{Machine Learning based Prediction Models}
\label{sec:ml-models}

We use a Gradient Boosting regressor (GBR) both as our prediction model, and
for assigning importances to input features. GBRs utilize an ensemble method
that assumes that the true regression function is a linear combinations of
several different base learners~\cite{freund2003efficient, friedman2001greedy}.
These base learners typically constitute decision trees, which have the benefit
of higher levels of human interpretability particularly in determining the
importances of input features.  Since the number of samples per application
dataset is relatively small, we solely consider GBR in our experiments to
determine feature importances. 

In addition to using GBR, we also train a neural network when combining
multiple datasets to create application-agnostic models in
Section~\ref{sec:agnostic}.  The larger combined dataset allows for more
complex models. We utilize a small neural network consisting of two 8-node
hidden layers each with a 50\% dropout connected to a final output layer with a
linear activation. The dropout layers randomly select 50\% of the nodes to
exclude from a layer during training and help to reduce
overfitting~\cite{srivastava2013improving}. GBRs can struggle with
extrapolation so a neural network was selected to counter this
specificity~\cite{prettenhofer2014gradient}.

\subsection{Training the Models}

In Sections~\ref{sec:model1} and~\ref{sec:model2}, we train separate machine
learning models for each of the first four datasets in Table~\ref{tab:apps}. We
perform a 20-fold cross-validation, where the dataset is split into 20 parts
randomly. One part is reserved for testing and the other 19 parts are used for
training. The inputs to the ML algorithms for creating the dataset-specific ML
models are: (1)~for each sample (job) in the training set, values of the
counters described in Table~\ref{tab:counters-derived} for the five minutes
prior to the start of that job are provided as the input features, and
(2)~execution time of each sample (job) is provided as the dependent variable
to be modeled.  Given a set of samples in the testing set, the model outputs
the predicted execution time of each testing sample based on the values of the
independent features (counter values) for that sample. We standardize all input
features (counter values) and the execution time of the training data (yielding
a mean of zero and standard deviation of one for each feature and the output
vector).  When testing a model, we apply the same standardization vectors
obtained from the training data on the test data.

In Section~\ref{sec:agnostic}, we create an application-agnostic model that can
predict the performance of an arbitrary application not included in the
training dataset. We create multiple training datasets by combining data samples from
different rows of Table~\ref{tab:apps}, and leaving some samples for testing
entirely. For example, in one instance, we combine the following three datasets
-- AMG 128, AMG 512, and MILC 128, train a model using the combined data, and
use the trained model to predict the performance of MILC 512. In a separate
study, we combine datasets by application type. For example, we combine all AMG
and MILC datasets, train a model, and use miniVite as an unseen testing
dataset.  Note that when we combine multiple datasets, we standardize their
features and execution times separately. We de-normalize the output by using a
standardization vector created from the oracle execution times of the testing
data.  If we were to use such a model for a new application, we would not have
a standardization vector for de-normalizing the predicted values. However, the
standardized output still allows analyzing relative expected performance, and
provides a strong heuristic for overall runtime with only an estimation of the
true application runtime distribution. 

\begin{figure*}[t]
  \centering
  \includegraphics[width=0.49\textwidth]{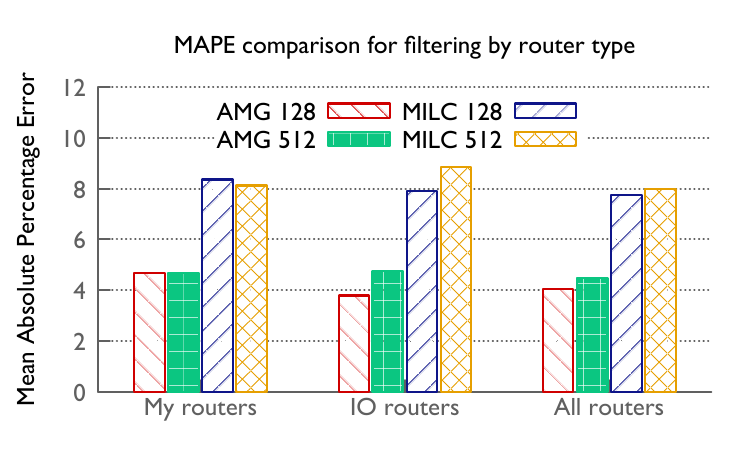}
  \hfill
  \includegraphics[width=0.49\textwidth]{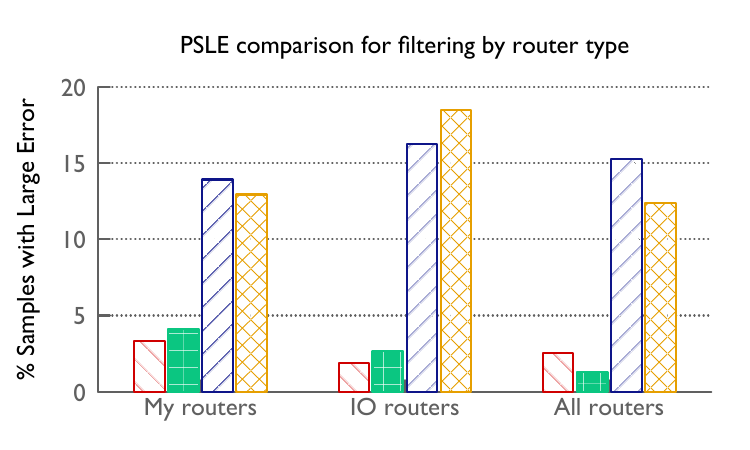}
  \caption{MAPE and PSLE scores for the GBR model when using different router types for filtering the system-wide data.}
  \label{fig:exp1-err}
\end{figure*}

\subsection{Calculating Feature Importances}

To evaluate the relative importances of the derived system-wide counters in
forecasting job runtime, we use the technique of recursive feature elimination
(RFE). For each application dataset, we perform a summary across a particular
router type using an aggregation function, and train a GBR model.  scikit
outputs importances of each feature, where the importance is computed as the
(normalized) total reduction of the mean squared error brought by that feature.
It is also known as the Gini importance.  We identify the worst performing
feature based on feature importances, drop that feature from the training data,
and train again with the smaller set of features. This process continues until
all features are eliminated. Finally, we create a ranking based on when each feature
is eliminated, and use this ranking to select the five best features and compute a relevance score
for each.

\begin{figure*}[t]
  \centering
  \includegraphics[width=0.49\textwidth]{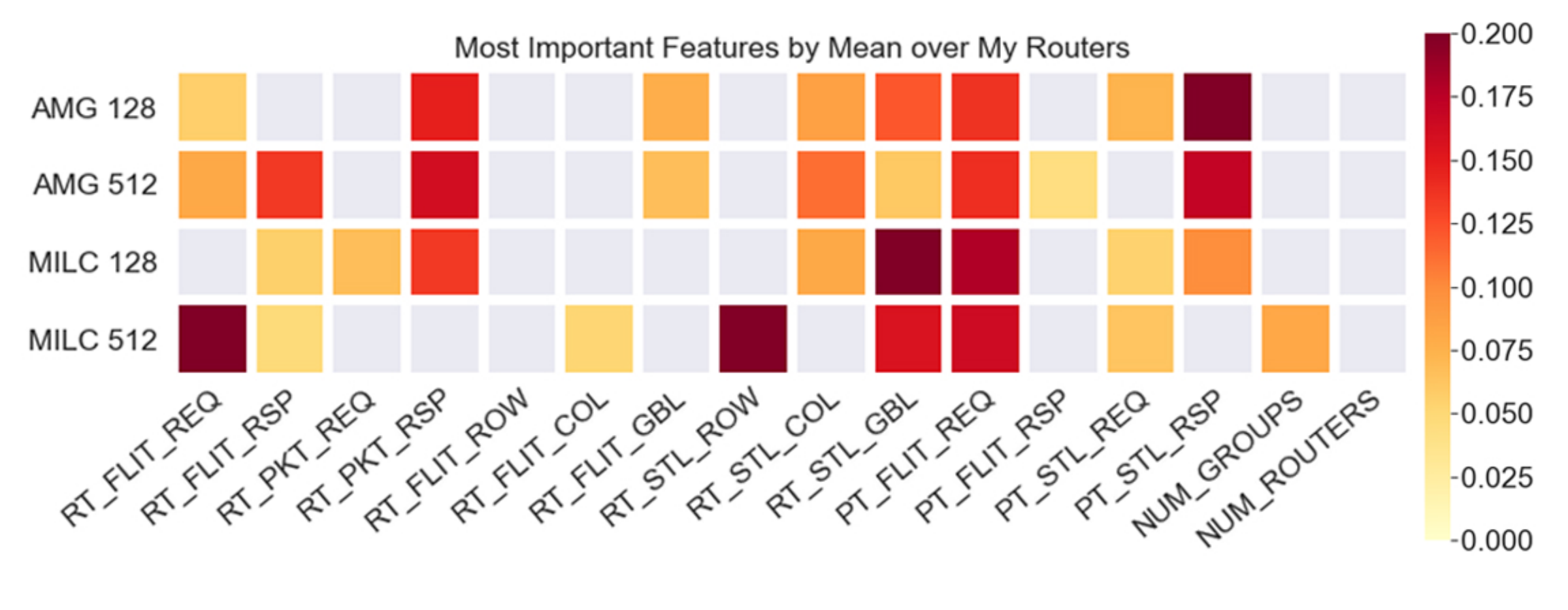}
  \hfill
  \includegraphics[width=0.49\textwidth]{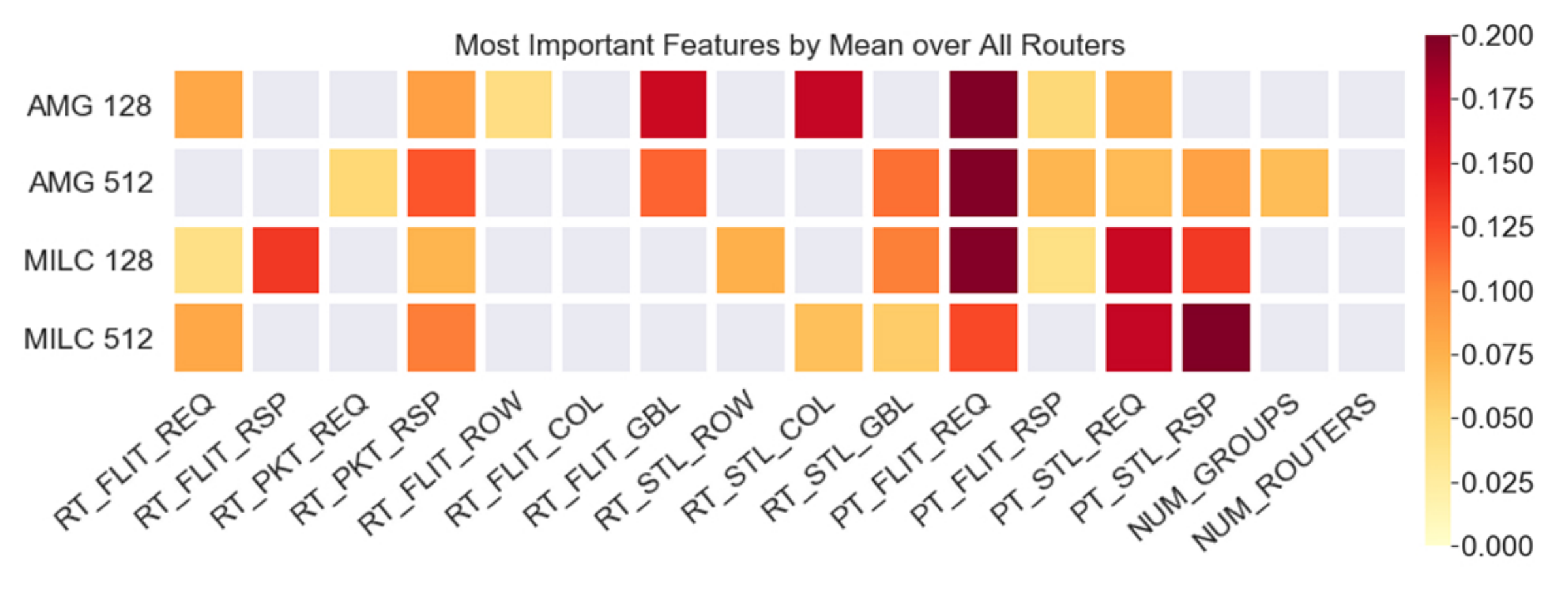}
  \caption{Relative importances of the most important counters obtained using RFE for different datasets (aggregation function: mean).}
  \label{fig:exp1-imp}
\end{figure*} 

We use a tree-based feature importance metric for the GBR. For each feature
utilized by each of GBR's decision trees, the total reduction in mean squared
error that can be attributed to a branch utilizing that feature is calculated
for all features~\cite{sandri2010analysis}.  Note that all input features in
our dataset are strictly numeric and thus less susceptible to biases sometimes
present in tree-based feature importances.  For each dataset, we perform this
calculation of feature importances 5-fold and average the results across all validation splits.

\subsection{Metrics for Evaluation}

We define two metrics that are indicative of the quality of predictions
w.r.t.~overall job runtime.  The first, mean absolute percentage error (MAPE),
calculates the mean of percentage errors observed for each test sample as
follows,
\[ \mathrm{MAPE} = \frac{1}{n} \sum_{i}^{n} \frac{|y_i-\hat{y}_i|}{y_i} \] 
where, $y_i$ is the true value, $\hat{y}_i$ is the predicted value, and $n$ is
the number of samples in the dataset.

We also define an additional metric to measure the relative accuracy of our
regression models: percent of samples with large error (PSLE). We define a test
sample to have a large error if the predicted value is more than $x$\% higher
or lower than its true value. For a dataset, PSLE is defined as,
\[ \mathrm{PSLE} = \frac{1}{n} \sum_{i=1}^{n} \mathit{LE}_{i} \]
\[  \mathit{LE}_{i} =
    \begin{cases}
        1,& \text{if } \frac{|y_i-\hat{y}_i|}{y_i} > x \\
        0,& \text{otherwise}
    \end{cases}
\]
We use $x = 0.15$ for the evaluation in this paper. This metric is important
because when a system wants to use the model, it may not need exact predictions
of the job runtime. It may be more important to predict the general trend i.e.
is the next job going to run reasonably fast or unreasonably slow?  We chose $x
= 0.15$ because if the ML model is mis-predicting by more than 15\%, then the
job scheduler should probably not use the results to decide if the job will run
slow or fast. We use MAPE and PLSE as opposed to the standard ML metrics because they provide
a more intuitive understanding of the efficacy of the model for the HPC
performance domain.

\section{Evaluation of Trained ML Models}
\label{sec:results}
We now evaluate and compare the prediction models trained on different datasets
and their combinations, for different router groupings and aggregation
functions.

\begin{figure*}[t]
  \centering
  \includegraphics[width=0.49\textwidth]{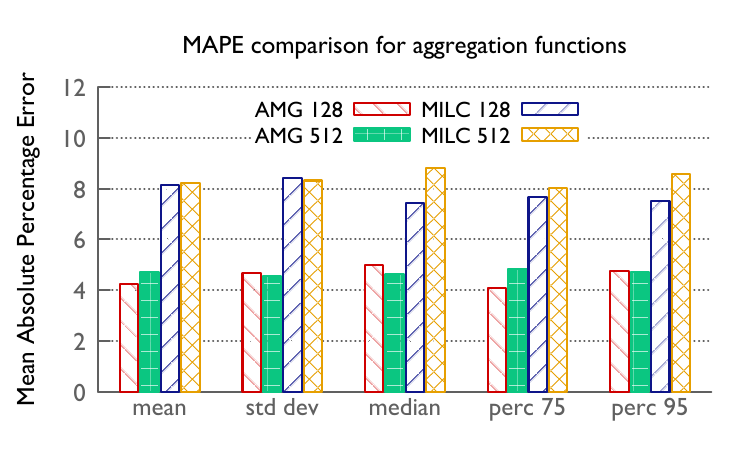}
  \hfill
  \includegraphics[width=0.49\textwidth]{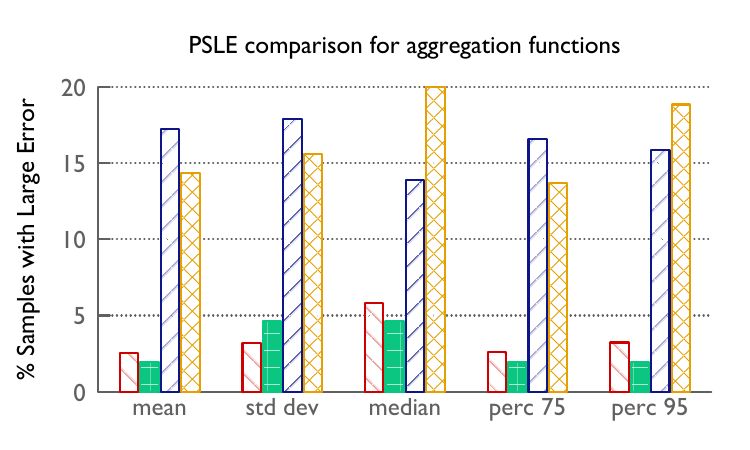}
  \caption{MAPE and PSLE scores for the GBR model when using different aggregation functions over all compute routers.}
  \label{fig:exp2-err}
\end{figure*}

\subsection{Models based on Different Router Groups}
\label{sec:model1}

In order to understand the significance of different router groups in
predicting performance, we created GBR models for each application dataset
using three different router groups -- my routers, IO routers, and all compute
routers. We observe that despite a small number of samples per application
dataset, all models perform sufficiently well (Figure~\ref{fig:exp1-err}). All
router groups are good at predicting AMG performance with MAPE below 5\% and
for MILC, which has a much higher variance in runtime, all MAPEs are below
10\%. The All routers grouping is somewhat better than the other two. We
see a similar trend with the Percentage Samples with Large Error (PSLE) metric
–– for AMG, the models predict worse than 15\% for less than 5\% of the test
data. However, for MILC, PSLE is somewhat higher due to the high variability in
the MILC dataset.  We also observe that we obtain good scores for both My
routers and All routers groupings. This indicates that both system
administrators and individual HPC users could achieve similar success in
predicting complex job execution with small amount of temporal information.

Using recursive feature elimination for both AMG and MILC datasets, we
calculate relative importance scores for all input features in the My routers
and All routers groupings. Figure~\ref{fig:exp1-imp} visualizes the importances
of the derived features for the four datasets (each row is for one dataset). We observe that the two router
groupings rely on some common features -- {\tt RT\_PKT\_RSP} (stalls on router
tiles), {\tt RT\_STL\_GBL} (stalls on global links), {\tt PT\_FLIT\_REQ}
(processor tile flits), and {\tt PT\_STL\_RSP} (processor tile stalls). It is
not surprising that job performance depends on stalls on router and processor
tiles, specifically on stalls on global links.  For both applications,
performance also depends on the number of flits sent on processor tiles.

\subsection{Models by Different Types of Aggregation}
\label{sec:model2}

In the previous section, we used the mean function to aggregate data over all
routers in a particular group.  We now consider several other aggregation
strategies, and the top performing aggregations for the all routers grouping
are shown in Figure~\ref{fig:exp2-err}. We observe that across these top
performing aggregation schemes, the prediction scores do not vary significantly
for the different applications. In terms of the application of these prediction
models in a system-wide job scheduler, we see this as a promising result as the
true mean across all routers is not needed. It highlights the potential for
strong results using a computationally less-expensive aggregation and for
accurate estimation using a small subset of routers.   

\begin{figure}[h]
  \centering
  \includegraphics[width=0.49\textwidth]{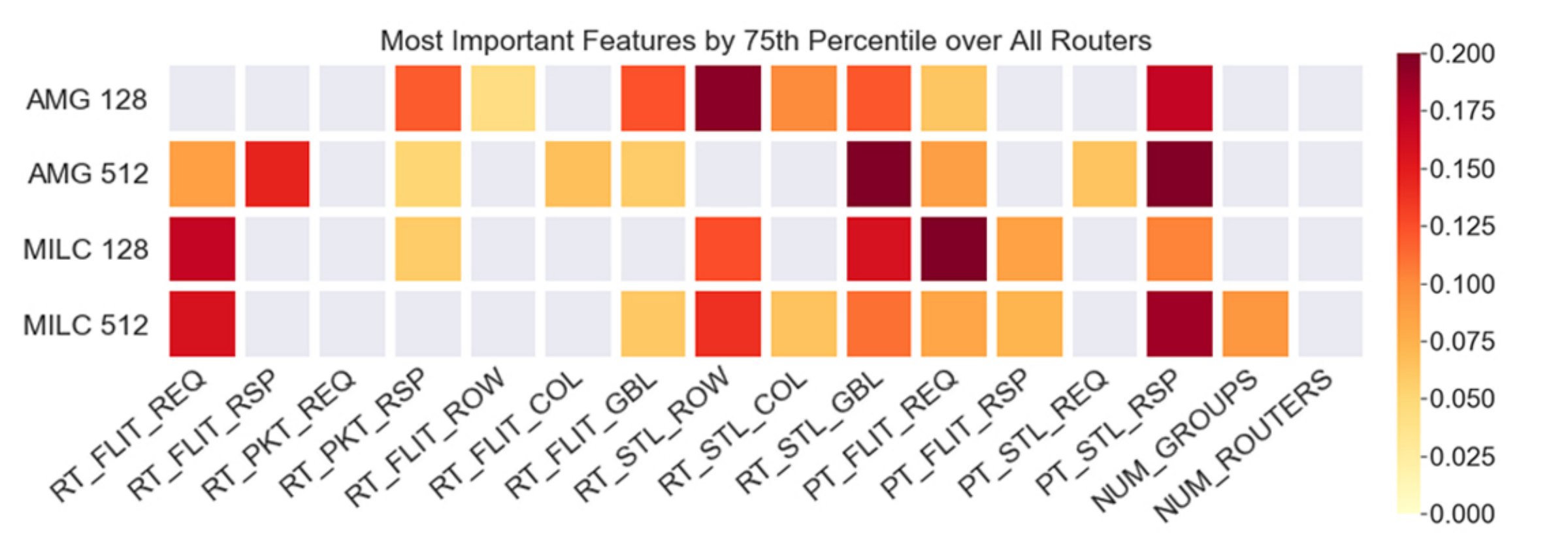}
  \caption{Relative importances of the most important counters obtained using RFE for different datasets (router grouping: all routers).}
  \label{fig:exp2-imp}
\end{figure}

Next, we perform RFE on the derived features when using the 75th percentile
function for aggregation. Figure~\ref{fig:exp2-imp} shows the feature
importances for the 75th percentile aggregation over the All routers grouping.
We see a consistent story in the feature importances to the mean aggregation
(Figure~\ref{fig:exp1-imp}) suggesting a robustness in this ranking strategy.
Some of the same derived counters appear to be the most important -- {\tt
RT\_STL\_GBL} (stalls on global links), {\tt PT\_FLIT\_REQ} (processor tile
flits), and {\tt PT\_STL\_RSP} (processor tile stalls).
{\tt RT\_PKT\_RSP} (stalls on router tiles) appears to be less important now
but some other features also appear to be important in this scenario: {\tt
RT\_FLIT\_REQ} (router tile flits), {\tt RT\_STL\_ROW} (stalls on row links),
and {\tt PT\_STL\_RSP} (processor tile stalls).

\begin{figure*}[t]
  \centering
  \includegraphics[width=0.49\textwidth]{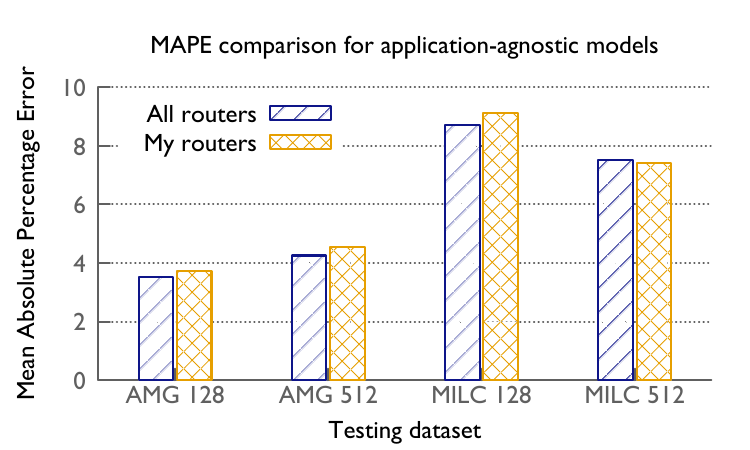}
  \hfill
  \includegraphics[width=0.49\textwidth]{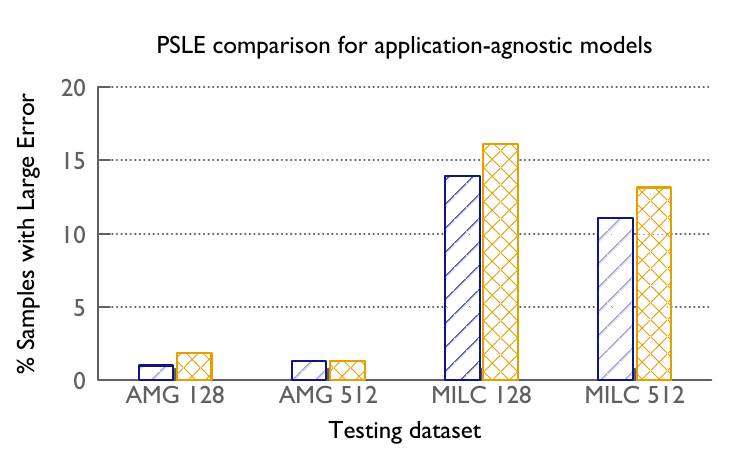}
  \caption{MAPE and PSLE scores for the NN model when using three datasets
for training and a fourth disjoint dataset for testing (x-axis label.) The
training dataset for each cluster is the combination of AMG 128, AMG 512, MILC
128, MILC 512 minus the dataset in the x-axis label.}
  \label{fig:exp3-1-err}
\end{figure*}

\begin{figure*}[t]
  \centering
  \includegraphics[width=0.49\textwidth]{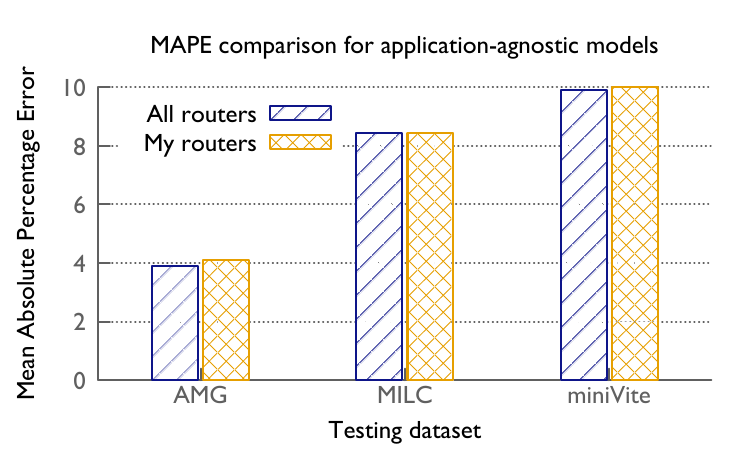}
  \hfill
  \includegraphics[width=0.49\textwidth]{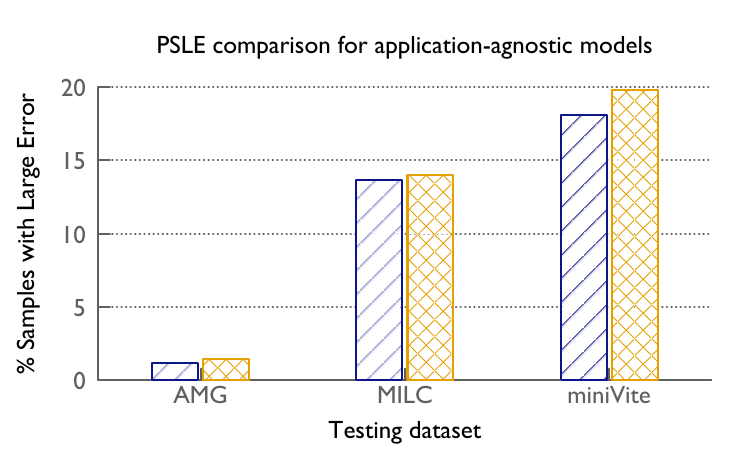}
  \caption{MAPE and PSLE scores for the NN model when combining datasets by
application type. Two applications are used for training and the third
application is used for testing (x-axis label.)}
  \label{fig:exp3-2-err}
\end{figure*}

\subsection{Application-agnostic Models}
\label{sec:agnostic}

Finally, we analyze the generalizability of the ML algorithms and their
prediction models.  Generalization refers to a model's ability to adapt to new
unseen data. When we reserve an entire application for testing and do not
include it in training, and repeat this with different applications, it tests
the generalizability of the model.  The end goal is to train a single model
that can accurately predict the standardized runtime for any application even
if we do not have data for that application in the training dataset. In the
first study of generalizability, we use four datasets -- AMG 128, AMG 512, MILC
128, and MILC 512. In turns, we use three of these datasets for training and
reserve the fourth dataset entirely for testing.  We segment the training data
into 8-fold cross-validation segments and train a model that can predict the
standardized runtime of any job in the testing set given the previous five
minutes of system data. Each prediction is later de-normalized with respect to
its application and an error metric is calculated for that prediction. For all
the results in this section, we use the neural network model, apply the mean
function to aggregate the data, and compare using the All routers versus My
routers data.


Figure~\ref{fig:exp3-1-err} shows the success of the trained models in terms of
their MAPE and PSLE scores. Comparing with Figure~\ref{fig:exp2-err}, we
observe that when multiple datasets are combined for training, the models
perform better in terms of predicting the execution times, compared to training
on a portion of the individual datasets by themselves. The MAPE for predicting
AMG 128 reduces from 4.23 to 3.61 and that for AMG 512 from 4.71 to 4.25.
Similarly the MAPE for predicting MILC 512 reduces from 8.21 when used by
itself for training to 7.5 when the other three datasets are combined for
training a model.  This improvement is likely due to the larger training
dataset (\tweakedsim 450 samples versus \tweakedsim 150) allowing for more
robust training of models.  We see this as a promising sign for future models
which could include tens of applications with hundreds of samples each and
likely even stronger and more generalizable predictions.  We also observe that
using the data from only the routers allocated to a job does not degrade the
models significantly. This suggests that in absence of system-wide data, an end
user can use data from the routers that they have access to via
their jobs.

In the second study of generalizability, we combine datasets by application
type and reserve one of the applications as unseen data for testing. For
example, when we combine all AMG and MILC datasets for training, we use the
miniVite dataset for testing.  Figure~\ref{fig:exp3-2-err} shows how these
application-agnostic models perform in terms of predicting the performance of
an unseen application. We observe that AMG has the lowest errors, followed by
MILC and then miniVite. On average, AMG has the lowest percentage of
communication with respect to its total execution tine, followed by MILC, and
then miniVite. As a result, AMG has the lowest performance variability and
miniVite the highest. We believe that this is the reason for the models
performing better in predicting AMG's performance as opposed to that of MILC
and miniVite. Nevertheless, the results are still encouraging. Even without any
data for an application being included in the training dataset, the ML models
exhibit reasonable success in performance prediction. We expect that adding
more applications with different computation and communication signatures to
our training dataset will improve the prediction scores for unseen
applications.

\begin{figure}[h]
  \centering
  \includegraphics[width=0.99\columnwidth]{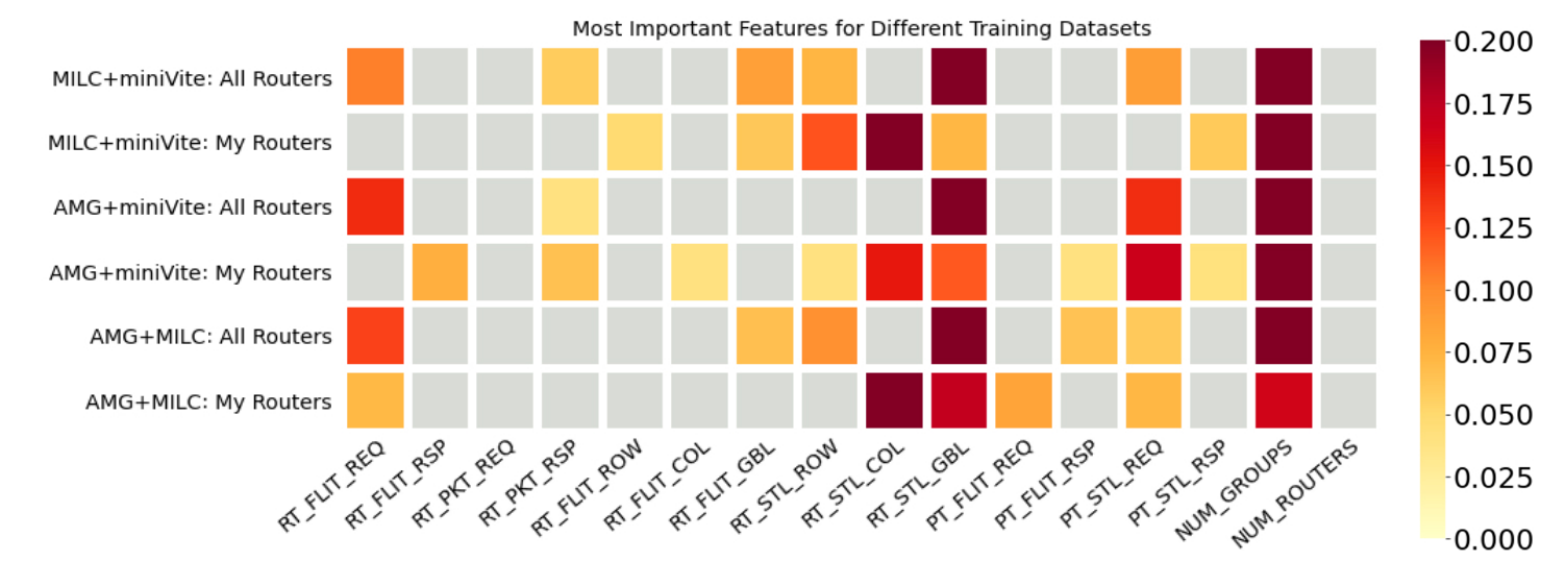}
  \caption{Relative importances of the most important counters obtained using RFE for different router groups in the application-agnostic model.}
  \label{fig:exp3-imp}
\end{figure}

Finally, we analyze feature importances when training the neural network model
using the combined datasets. Figure~\ref{fig:exp3-imp} shows the relative
feature importances for three different training datasets (AMG+MILC,
AMG+miniVite, and MILC+miniVite), and two filterings (All routers and My
routers). Surprisingly, {\tt NUM\_GROUPS} emerges as a highly important
feature. In theory, one would expect that the placement of a job should have
little impact on its performance due to adaptive indirect (UGAL)
routing~\cite{ugal}. However, in practice, it is possible that when a job is
spread over more groups, the likelihood of encountering congestion on global
links increases. {\tt RT\_STL\_GBL} (stalls on global links) is also important
for predicting all three applications as we had observed in the previous plots.
We also observe that while applications share common important features, some
features are only important for certain datasets. We notice that {\tt
PT\_STL\_REQ} (processor request stalls) is more important when training using
the AMG+miniVite dataset. A feature that is important when filtering by My
routers but not All routers is {\tt RT\_STL\_COL} (stalls on black links). On
the other hand, {\tt RT\_FLIT\_REQ} (router request flits) is
important when filtering by All routers.

\section{Influencing Job Scheduling Decisions}
\label{sec:impl}
We now discuss how the findings in previous sections could be used by a job
scheduler or HPC user for labeling incoming jobs in the queue as likely to run
relatively fast or slow. The hypothesis is that by selecting a small number of
features (network counters) based on feature importances, and analyzing their
values when a new job is ready to be scheduled, the job scheduler can quickly
determine if the job will run slow or fast.  If this succeeds, a job scheduler
can decide to monitor certain features continuously, based on feature
importances derived from the application-agnostic models in
Section~\ref{sec:agnostic}.

We select the three most important features from the application-agnostic
models in Figure~\ref{fig:exp3-imp}: {\tt NUM\_GROUPS}, {\tt RT\_STL\_GBL}, and
{\tt RT\_STL\_COL}. Next, we classify samples (jobs) in three of our datasets
(AMG 512, MILC 512, and miniVite 128) as ``likely fast'' or ``likely slow''
based on whether the system-wide values of these three counters in the five
minutes prior to that job running were below or above the median of all
observed values respectively. Once the jobs in a dataset have been classified
into likely fast or slow based on the values of the selected network counters,
we analyze their actual execution times (ground truth) to check if our
classification is statistically significant.

\begin{figure}[h]
  \centering
  \includegraphics[width=0.49\textwidth]{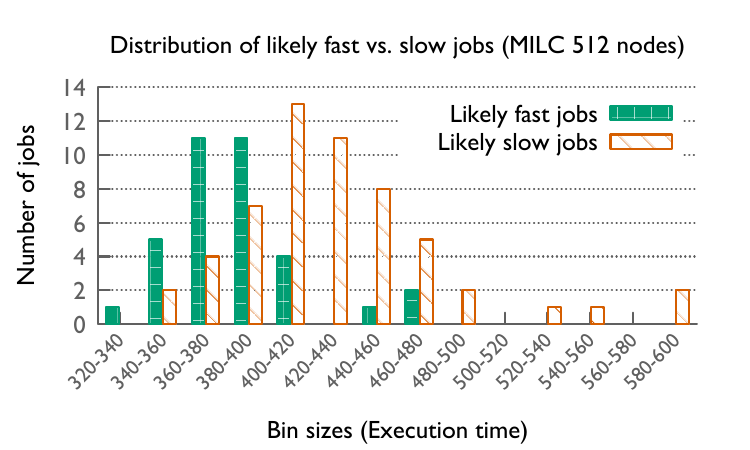}
  \caption{Distribution of actual runtimes of likely fast versus slow jobs 
of MILC when considering above median values of three features: RT\_STL\_COL,
RT\_STL\_GBL, and NUM\_GROUPS}.
  \label{fig:hist}
\end{figure}

Figure~\ref{fig:hist} shows the distributions of the execution times of the
likely fast and slow sets of jobs in the MILC 512 dataset. The histograms were
generated with a fixed number of bins over the entire execution time range of a
given dataset. Given the right skew present in application runtimes, we elected
to use the Kruskal-Wallace H.~Test to test for a difference in medians between
the runtimes of likely fast and slow jobs in each application dataset. We found
that for all the applications, the calculated p-value was below 3e-05,
indicating a statistically significant difference between the likely fast and
slow execution times. Note that a one-way ANOVA test yields statistically
significant results below the 1\% threshold for all applications.
Table~\ref{tab:mean} compares the mean execution times of the likely fast
versus slow jobs in each dataset. We can see that the difference is
significant, especially for MILC and miniVite in spite of their predictions
being poorer than AMG.  This is a powerful result with significant implications
for improving application performance and reducing variability.

\begin{table}[h]
\centering
\caption{Mean execution times (in seconds) of the likely fast and slow subsets of jobs in each dataset}
\begin{tabular}{lrr} \toprule
Application  & Fast jobs & Slow jobs \\ \midrule
AMG 128      & 260.12 & 274.24 \\
AMG 512      & 379.93 & 410.68 \\
MILC 128     & 317.31 & 398.43 \\
MILC 512     & 389.61 & 445.80 \\
miniVite 128 & 309.96 & 372.26 \\
\bottomrule
\end{tabular}
\label{tab:mean}
\end{table}

We foresee two immediate applications of such a system.  Many HPC systems set
an upper bound on the time requested for a job (12 or 24 hours). As a result,
science simulations requiring months of running time must be submitted as
multiple jobs to the job queue periodically. While the inputs may change
somewhat as computation progresses, the overall runtime profile and networking
requirements of an application usually do not vary significantly, providing an
optimal use case for our prediction system.  When a job gets scheduled,
individual HPC users can gather network counter data for a few minutes on all
the routers assigned to their job and use our pre-trained models to predict the
expected runtime of their application. They can use this prediction to decide
whether to go ahead with launching their application or to give the
allocation back and request another job later when the system is less congested.

Second, these results suggest that an intelligent job scheduler can monitor a
handful of counters, and use their current values to determine if, for example,
communication-heavy jobs will perform well or poorly if scheduled right away.
Figure~\ref{fig:hist} demonstrates the power of predicting
job execution time solely based on the median aggregation of just three network
counters. While in this paper, we analyzed these jobs after they had run, a job
scheduler would have access to similar counter data through LDMS at the time
when a job is ready to be scheduled.  Such adaptive decisions based on
light-weight monitoring of a few hardware counters on a subset of routers could
prove to be immensely useful.

\section{Summary and Future Work}
In summary, we presented a data analytics study of longitudinal system-wide
monitoring data to predict the performance of unseen jobs pending in the
scheduler queue. We presented a pipeline for extracting relevant data from a
time series, creating interpretable derived features, reducing the data, and
filtering and aggregating it in meaningful ways. We then created several
prediction models that only look at prior system state before a job starts
executing to predict its runtime. Our models performed well on two different
metrics and also helped in identifying important input features.  We then
demonstrated the use of three network hardware counters to classify jobs in a
dataset into likely fast and likely slow with statistically significant
results. This demonstrates that an intelligent job scheduler could use a
similar simple mechanism to forecast the performance of a pending job. Our
proposed pipeline and analysis remain relevant across different HPC systems and
routers as our work can be applied to any system with LDMS data collection.

In the future, we plan to perform more detailed analysis of system- and
facility-wide monitoring data to detect patterns and anomalies in it. We also
plan to create a large library of performance datasets that can be used to
train machine learning models that would be successful in predicting the
performance of any new code. We also plan to modify existing job scheduling
algorithms to incorporate machine learning models for making real-time
decisions that reduce performance variability.

\section*{Acknowledgment}
This material is based upon work supported in part by the National Science
Foundation under Grant No.~2047120.

\bibliographystyle{IEEEtran}
\bibliography{./bib/pssg,./bib/cite}

\end{document}